\documentclass[12pt]{article}
\usepackage{amssymb,amsmath,epsfig}
\allowdisplaybreaks

\begin{document}
\title{\bf Cylindrical Thin-shell Wormholes in $f(R)$ gravity}

\author{M. Sharif \thanks{msharif.math@pu.edu.pk} and Z. Yousaf
\thanks{z.yousaf.math@live.com}\\
Department of Mathematics, University of the Punjab,\\
Quaid-e-Azam Campus, Lahore-54590, Pakistan.}
\date{}

\maketitle
\begin{abstract}
In this paper, we employ cut and paste scheme to construct
thin-shell wormhole of a charged black string with $f(R)$ terms. We
consider $f(R)$ model as an exotic matter source at
wormhole throat. The stability of the respective solutions are
analyzed under radial perturbations in the context of
$R+{\delta}R^2$ model. It is concluded that both stable as well as
unstable solutions do exist for different values of $\delta$. In the
limit $\delta{\rightarrow}0$, all our results reduce to general
relativity.
\end{abstract}

\section{Introduction}

Wormhole (WH) is a hypothetical tunnel, path or bridge associating
two different portions of the spacetime under which observers may
pass freely. Flamm (1916) was the first who found Schwarzschild
solution as non-traversable WH while Einstein and Rosen (1935)
investigated WH solutions with event horizon. Morris and Thorne
(1988) claimed that a WH can be made traversable if it is supported
by exotic matter. The existence of exotic matter at the WH throat
made it a burning issue which attracted many researchers. It is
interesting to mention that stability phases of the self-gravitating
bodies lead to different evolutionary processes in the universe. In
this context, the instability investigation for the collapsing
processes has been widely performed (Herrera et al. 1989, Chan et
al. 1993, 1994, Herrera and Santos 1997, Herrera et al. 2012,
Pinheiro and Chan 2013). Moreover, the stability analysis of WHs
against small perturbations is also a core issue in astrophysics.

It is argued that exotic matter requirement in WHs can be refrained
in modified theories of gravity (Gravanis and Willison 2007,
Garraffo et al. 2008, Anchordoqui et al. 1997). Thin-shell WHs are
built up by cut and paste scheme from black holes. In this
technique, the exotic matter required to construct WH is settled at
the shell and the matching condition is used for its analysis. The
surface stresses in this framework are computed by using the
Darmois-Israel formalism (Israel 1966, 1967, Papapetrou and Hamoui
1968). One can investigate the dynamical stability of thin-shell WHs
either by analyzing a linearized stability procedure about a static
solution (Poisson and Visser 1995, Lobo 2006), or by considering a
particular equation of state (EoS) (Visser 1990, Kim 1992, Kim et
al. 1993). The stability of this sort of matter distribution is
being analyzed in general relativity (Barcel and Visser 2000) as
well as in extended gravity theories (Anchordoqui 1998, Eiroa and
Simeone 2005).

Brady et al. (1991) studied the dynamics of an infinitely thin
massive shell and concluded that such stable shell has relatively
larger surface energy density than pressure. Cl\'{e}ment (1995)
presented multi WH solutions in which the spacetime asymptotically
inclines to the conical cosmic string spacetime. Aros and Zamorano
(1997) found a solution which may be regarded as a traversable
cylindrical WH within the global cosmic string core. Eiroa and
Romero (2004) extended their results by invoking electric charge
while Lobo and Crawford (2004) generalized this analysis with
cosmological constant. Eiroa and Simeone (2004) discussed the
dynamics of thin-shell WHs under non-rotating cylindrical
background. The same authors (2005) extended this work for charged
Lorentzian WHs in the framework of dilaton gravity and calculated
the total quantity of exotic matter.

Thibeault et al. (2006) investigated $5D$ thin-shell WHs in the
scenario of modified theory. Rahaman et al. (2007) constructed
thin-shell WH in the scenario of heterotic string theory and
investigated its stability against perturbation. Eiroa and Simeone
(2007) used Chaplygin EoS to study the stability of thin-shell WHs
by introducing a new scheme. They applied this approach to analyze
the stability of WHs constructed from the Schwarzschild,
Schwarzschild de Sitter, Schwarzschild anti-de Sitter and
Reissner-Nordstr\"{o}m spacetimes. Sharif and Azam (2013)evaluated
unstable and stable distributions of thin-shell in cylindrical
symmetry.

Nojiri et al. (1999a, 1999b) found some induced WH solutions
incorporating increasing red-shift function and throat radius for
some specific values of initial conditions. Nojiri and Odintsov
(2007) described late-time (quintessence/ phantom) universe filled
with dark sources arising from modified gravity theories with
different choices of generic functions of $f(R)$ and $f(R,G)$.
Nojiri and Odintsov (2011) also presented various aspects of $f(R)$
gravity and claimed that there is a variety of $f(R)$ models that
are well-consistent with local tests and observational data
(Capozziello et al. 2009, 2010, Dev et al. 2008). Bamba et al.
(2012) reviewed different dark energy cosmological models which may
lead to the accelerating expansion of the universe.

Since exotic matter does not satisfy null energy condition, so
researchers are interested to find realistic sources that can
support WHs. Furey and DeBenedicits (2005) studied WH throats for
$R^{-1}$ and $R^2$ gravity and concluded that static WHs respect
null energy condition. DeBenedicts and Horvat (2012) extended these
results for a model of the form $f(R)=\sum\alpha_nR^n$. Lobo and
Oliveria (2009) obtained static WH solutions for traceless matter by
choosing barotropic EoS in $f(R)$ gravity.

In this paper, we investigate the role of charge in the stability of
thin-shell WH by cut and paste technique with $f(R)$ terms. In order
to check the dynamical stability, we choose Darmois-Israel matching
conditions. The paper is planned as follows. In
section \textbf{2}, we obtain general formulation required for the
study of thin-shell WH. Section \textbf{3} is devoted to analyze the
linearized stability of thin-shell WHs while in section \textbf{4},
we apply this formalism on charged black string with $f(R)$ terms.
In the last section, we conclude our results.

\section{Thin-Shell Wormhole and $f(R)$ Gravity}

The modified form of Einstein-Hilbert action in $f(R)$ gravity can
be written as
\begin{equation*}\label{1a}
\mathcal{A}_{f(R)}=\frac{1}{2\kappa}\int
d^4x\sqrt{-g}f(R)+\mathcal{A}_M,
\end{equation*}
where $\mathcal{A}_M$ and $f(R)$ are the matter action and a
non-linear real function of the curvature $R$, respectively. The
field equations are evaluated by giving variation in the above
action with respect to $g_{\alpha\beta}$ as follows
\begin{equation*}\label{1b}
f_RR_{\alpha\beta}-\frac{1}{2}fg_{\alpha\beta}-\nabla_{\alpha}
\nabla_{\beta}f_R+g_{\alpha\beta}{\Box}f_R=\kappa S_{\alpha\beta},
\end{equation*}
where $S_{\alpha\beta}$ is the energy-momentum tensor,
$\Box=\nabla^\alpha\nabla_{\alpha}$, $\nabla_\alpha$ is the
covariant derivative and $f_R=\frac{df}{dR}$. This equation can be
formulated alternatively in the form of general relativity (GR)
field equations as
\begin{equation*}\label{1c}
G_{\alpha\beta}=\frac{\kappa}{f_R}(S_{\alpha\beta}+\overset{(D)}{S_{\alpha\beta}}),
\end{equation*}
with
\begin{equation*}\label{1d}
\overset{(D)}{S_{\alpha\beta}}=\frac{1}{\kappa}\left\{\frac{f-Rf_R}{2}
g_{\alpha\beta}+\nabla_{\alpha}\nabla_{\beta}f_R-\Box
f_Rg_{\alpha\beta}\right\}.
\end{equation*}
In $f(R)$ gravity, the junction conditions over a timelike boundary
surface $\Sigma$ in $4D$ manifold can be found by projecting the
above equations on the boundary surface $\Sigma$.

The extrinsic curvature linked with two portions of the hypersurface
$\Sigma$ is
\begin{equation}\label{1}
K^{\pm}_{ij}=-n^{\pm}_{\sigma}\left.\left(\frac{{\partial}^2x^{\sigma}_{\pm}}
{{\partial}{\zeta}^i{\partial}{\zeta}^j}+{\Gamma}^{\sigma}_{{\alpha}{\beta}}
\frac{{{\partial}x^{\alpha}_{\pm}}{{\partial}x^{\beta}_{\pm}}}
{{\partial}{\zeta}^i{\partial}{\zeta}^j}\right)\right|_{\Sigma},
\end{equation}
where ${\zeta}^j$, $x^{\sigma}$ and  $\Gamma^\sigma_{\alpha\beta}$
are the coordinates of the hypersurface, the four dimensional
manifold components and connection components related with the
metric $g_{\alpha\beta}$ respectively, while
\begin{equation}\label{2}
n^{\pm}_{\sigma}=\left|g^{\alpha\beta}\frac{{\partial}f}{{\partial}
x^{\alpha}}\frac{{\partial}f}{{\partial}x^{\beta}}\right|,
\end{equation}
are the unit normals $(n_{\sigma}n^\sigma=1)$. Consequently, the
Lanczos equations (Musgrave and Lake 1996) with $f(R)$ terms
Capozziello and Laurentis 2011, Sharif and Yousaf 2013a, 2013c,
2013d, 2013e) take the form
\begin{equation}\label{3}
\frac{\kappa}{{\alpha}f_R}\left({\alpha}S^i_{~j}+\overset{~~~(D)}{S^i_{~j}}\right)
=-\left(k^i_j-\delta^i_jk^a_a\right),
\end{equation}
where $\alpha^2=\frac{\Lambda}{3},~(\Lambda$ is the cosmological
constant), $S^i_{~j}$ and $\overset{~~~(D)}{S^i_{~j}}$ are the
energy-momentum tensor of the usual and effective matter on the
hypersurface, respectively and $k_{ij}=K^+_{ij}-K^-_{ij}$. The GR
Lanczos equations (Musgrave and Lake 1996) can be recovered from the
above equation under the limit $f(R)\rightarrow R$.

We construct a thin-shell WH of static cylindrically metric whose
line element is of the form (Lemos and Zanchin 1996)
\begin{equation}\label{4}
ds^2=-G(r)dt^{2}+G^{-1}(r)dr^{2}+N(r)(d\phi^{2}+\alpha^2{dz^2}),
\end{equation}
where
\begin{equation}\label{3a}
G(r)=r^2{\alpha}^2-\frac{4M}{r\alpha}+\frac{4q^2}{r^2\alpha^2},
\quad N(r)=r^2,
\end{equation}
$q$ and $M$ are the charge density and ADM mass, respectively. The
outer and inner charged black string horizons are given by
\begin{equation}\label{8}
r_{h\pm}=\frac{4^{1/3}}{2}
\left[\sqrt{s}\pm\left\{2\sqrt{s^2-q^2\left(\frac{2}{M}\right)^{\frac{4}{3}}}
-s\right\}^\frac{1}{2}\right]\frac{M^{\frac{1}{3}}}{\alpha},
\end{equation}
where
\begin{equation}\label{9}
s=\left(\frac{1}{2}-\frac{1}{2}\sqrt{1-\frac{64q^6}{27M^4}}\right)^{1/3}
+\left(\frac{1}{2}-\frac{1}{2}\sqrt{1-\frac{64q^6}{27M^4}}\right)^{1/3}.
\end{equation}
It is worth mentioning here that the the given spacetime does not
possess event horizon for $q^2>\frac{3}{4}M^{4/3}$ implying that
Eq.(\ref{8}) is valid only if $q^2\leq\frac{3}{4}M^{4/3}$. For
$q^2=\frac{3}{4}M^{4/3}$, the outer and inner horizons merge into
each other, representing extremal black string. We take radius $a$
and choose two $4D$ copies $\mathcal{W}^-$ and $\mathcal{W}^+$ with
radius $r\geq a$ and paste them at the boundary surface $\Sigma$
defined by $r-a=0$, thus giving a geodesically complete new manifold
$\mathcal{W}=\mathcal{W}^-\cup\mathcal{W}^+$. If the geometry is let
to open at $\Sigma$, then this leads to a cylindrical thin-shell WH
with two parts associated by a throat at hypersurface (flair-out
condition). It is mentioned here that radius $a$ is chosen to be
greater than $r_h$ such that there are no singularities and horizons
in $\mathcal{W}$.

To investigate this traversable WH, we use the standard
Darmois-Israel formalism (Israel 1966, 1967, Papapetrou and Hamoui
1968). The wormhole throat is placed at the synchronous timelike
hypersurface with coordinates $\zeta=(\tau,~\phi,z)$ where $\tau$
represents proper time on the boundary surface. Using Eq.(\ref{1}),
we obtain
\begin{equation}\label{5}
K^{\pm}_{\tau\tau}=\mp\frac{2\ddot{a}+G'(a)}{2\sqrt{\dot{a}^2+G(a)}},
\quad K^{\pm}_{\phi\phi}= \pm
a\sqrt{\dot{a}^2+G(a)}=\frac{1}{\alpha^2}K^{\pm}_{zz}.
\end{equation}
The matter quantities $S^\tau_{~\tau}=-\sigma$ and
$S^\phi_{~\phi}=S^z_{~z}=P$ turn out to be
\begin{align}\nonumber
\sigma&=-\frac{4f_R}{a\kappa}\sqrt{\dot{a}^2+G(a)}+\frac{1}{{\alpha}\kappa}
\left\{\frac{f-Rf_R}{2}+G(a)f''_R+G(a)f'_R\left(\frac{N'(a)}{N(a)}
\right.\right.\\\label{6}
&+\left.\left.\frac{G'(a)}{2G(a)}\right)\right\},\\\nonumber
P&=\frac{f_R}{a\kappa}\left(\frac{2a\ddot{a}+2\dot{a}^2+2G(a)+aG'(a)}
{\sqrt{\dot{a}^2+G(a)}}\right)-\frac{1}{{\alpha}\kappa}
\left\{\frac{f-Rf_R}{2}+G(a)f''_R\right.\\\label{7}
&+\left.G(a)f'_R\left(\frac{N'(a)}{2N(a)}+\frac{G'(a)}{G(a)}\right)\right\}.
\end{align}

The stability of $f(R)$ models is also a significant issue which is
well discussed in the literature (Faraoni 2005, Faraoni and Nadeau
2005, Capozziello et al. 2004, 2006a, 2007, Capozziello et al.
2006). We take a familiar $f(R)$ model proposed by Starobinsky
(1980)
\begin{equation}\label{10}
f(R)=R+\delta R^{2},
\end{equation}
This model can explain the inflation period of the universe and is
stable for $\delta>0$ representing $f_{RR}>0$ (Noakes 1983; Sharif
and Yousaf 2013b). Besides substituting for dark energy at cluster
and stellar scales, $f(R)$ gravity can be used to present as an
alternate for dark matter (DM) (Capozziello et al. 2004, 2006a,
2006b, 2007). Thus the given $f(R)$ model was claimed both as DM
model with $\delta=\frac{1}{6M^2}$ (Cembranos 2009, 2011) and as an
inflationary prospect. For DM model, $M$ is figured out as
$2.7\times10^{-12}GeV$ with $\delta\leq2.3\times10^{22}Ge/V^2$
(Sotirou and Faraoni 2010). We are concentrated on this model to
investigate WH solutions in $f(R)$ gravity. Einstein theory is
recovered if $\delta=0$ thereby giving classically stable black
hole.

The accelerated expanding behavior of the universe triggered to
explore new matter that violates the strong energy condition called
dark energy. Pure Chaplygin gas obey EoS $P=-\frac{B}{\sigma}$
(Kamenshchik et al. 2001; Gorini et al. 2004), where $B>0$. Here we
are introducing this source just to solve the cumbersome set of
equations. Thus we have used its simplified version instead of
generalized Chaplygin gas EoS. Some authors (Hochberg et al. 1997,
Nojiri et al. 1999a, 1999b) presented numerical and analytical
spherically symmetric WH solution thus suggesting possibility of
inducing WHs at the early universe. Here, we also try to induce WH
solution at the early time universe (in the quantum era) with the
help of $R+{\delta}R^2$ model. It is well-known that if WHs are
studied in the early universe then quantum effects (Duff 1994) may
play significant anomaly effects. Using Eqs.(\ref{6}), (\ref{7}) and
(\ref{10}) in EoS, we obtain
\begin{equation}\label{11}
2a\ddot{a}+2\dot{a}^2+2G(a)+aG'(a)=\left[X_P
-\frac{Ba\kappa^2}{aX_{\sigma}-4f_R\sqrt{\dot{a}^2+G(a)}}\right]
\frac{a\sqrt{\dot{a}^2+G(a)}}{f_R},
\end{equation}
where
\begin{align*}\nonumber
&X_{\sigma}=\frac{\delta}{{\alpha}}\left[-\frac{R^2(a)}{2}+2G(a)R''(a)+2G(a)R'(a)
\left(\frac{N'(a)}{N(a)}+\frac{G'(a)}{2G(a)}\right)\right],\\\nonumber
&X_P=\frac{\delta}{{\alpha}}\left[-\frac{R^2(a)}{2}+2G(a)R''(a)+2G(a)R'(a)
\left(\frac{N'(a)}{2N(a)}+\frac{G'(a)}{G(a)}\right)\right],
\\\nonumber
&R(a)=G''(a)+2\frac{G''(a)N(a)}{G(a)}+\frac{G'(a)}{G(a)}\left(2G'(a)
-\frac{G(a)N'(a)}{2N(a)}\right),
\end{align*}
$X_\sigma$ and $X_P$ in Eq.(\ref{11}) represent $f(R)$ higher
curvature terms. This is the required differential equation that the
thin-shell WH (with throat radius $a$ supported by an exotic matter)
should satisfy. Using EoS, we can also have
\begin{align}\label{12}
P'=-P\frac{\sigma'}{\sigma}, \quad
\sigma'+2P'=\sigma'\left(1-\frac{2P}{\sigma}\right).
\end{align}
These relations will be helpful to eliminate $\sigma'$ as well as
$P'$ terms from the first and second derivatives of the potential
function.

\section{Stability Analysis}

In this section, we investigate the stability of static
configurations of the thin-shell WH framed within $f(R)$ gravity. In
this scenario, the surface pressure, energy density and dynamical
equation with static background yield
\begin{align}\label{13}
&P_0=\frac{(1+2{\delta}R_0)}{\kappa
a_0}\left(\frac{2G(a_0)+a_0G'(a_0)}{\sqrt{G(a_0)}}\right)
-\frac{X_{P0}}{\kappa},\\\label{14}
&\sigma_0=-\frac{4(1+2{\delta}R_0)}{\kappa
a_0}\sqrt{G(a_0)}+\frac{X_{\sigma0}}{\kappa},\\\label{15}
&2G(a_0)+aG'(a_0)=\frac{a_0\sqrt{G(a_0)}}{(1+2{\delta}R_0)}\left[X_{P0}
-\frac{Ba_0\kappa^2}{aX_{\sigma0}-4(1+2{\delta}R_0)\sqrt{G(a_0)}}\right],
\end{align}
where $X_{\sigma0},~X_{P0}$ and $R_0$ are evaluated at $a=a_0$. The
conservation equations help to examine many useful properties of the
WH throat such as variation of the throat internal energy and work
which internal forces in the throat has done. The energy density of
the surface and isotropic pressure obeying conservation equation can
be written as
\begin{align*}\nonumber
\frac{d}{d\tau}({\Delta}\sigma)+P\frac{d\Delta}{d\tau}=0,
\end{align*}
where $\Delta=4\pi a^2$, giving
\begin{align}\label{16}
\sigma'=-\frac{2}{a}(P+\sigma).
\end{align}

Equation of motion, about $a=a_0$, against radial perturbation
provides an efficient way to study the dynamics of thin-shell WHs.
Equations (\ref{6}) and (\ref{10}) lead to
\begin{equation*}\nonumber
\dot{a}^2+\Phi(a)=0,
\end{equation*}
where
\begin{align}\label{17}
\Phi(a)=G(a)-\frac{\kappa^2a^2}{16(1+2{\delta}R)^2}\left(\frac{X_\sigma}
{\kappa}-\sigma\right)^2,
\end{align}
is the the potential function whose first and second derivatives can
be found by using Eq.(\ref{12}) as
\begin{align}\nonumber
&\Phi'(a)=G'(a)-\frac{\kappa^2a^2}{8(1+2{\delta}R)^2}
\left(\frac{X_\sigma}{\kappa}-\sigma\right)
\left\{\frac{X'_\sigma}{\kappa}-\frac{1}{a}\left(2P
+\sigma-\frac{X_\sigma}{\kappa}\right)\right.\\\label{18}
&+\left.\frac{2{\delta}R'}{(1+2{\delta}R)^3}\left(
\frac{X_\sigma}{\kappa}-\sigma\right)\right\},\\\nonumber
&\Phi''(a)=G''(a)-\frac{\kappa^2a}{(1+2{\delta}R)^2}
\left[\frac{X'_\sigma}{\kappa}-\frac{1}{a}
\left(2P+\sigma-\frac{X_\sigma}{\kappa}\right)
+\frac{2{\delta}R'}{(1+2{\delta}R)^3}\right.\\\nonumber
&\times\left.\left(
\frac{X_\sigma}{\kappa}-\sigma\right)\right]\left[\left(
\frac{X_\sigma}{\kappa}-\sigma\right)\left(\frac{1}{4}
+\frac{18{\delta}R'}{(1+2{\delta}R)}\right)
+\frac{a}{8}\left\{\frac{X'_\sigma}{\kappa}+\frac{2}{a}
(\sigma+P)\right\}\right]\\\nonumber
&-\frac{\kappa^2a^2}{8(1+2{\delta}R)^2}
\left(\frac{X_\sigma}{\kappa}-\sigma\right)\left[
\frac{X''_\sigma}{\kappa}+\frac{1}{a^2}\left(2P
+\sigma-\frac{X_\sigma}{\kappa}\right)+\frac{1}{a}\left\{
\frac{2}{a}(P+\sigma)\right.\right.\\\nonumber
&\times\left.\left.\left(1-\frac{2P}
{\sigma}\right)+\frac{X'_\sigma}{\kappa}\right\}
+\frac{2\delta}{(1+2{\delta}R)^3}\left\{
\left(\frac{X_\sigma}{\kappa}-\sigma\right)\left(R''
-\frac{3R'^2}{(1+2{\delta}R)} \right)\right.\right.\\\label{19}
&\left.\left.+R'\left(\frac{X'_\sigma}{\kappa}
+\frac{2}{a}(P+\sigma)\right)\right\}\right].
\end{align}
Evaluating the above equation at $a=a_0$ and inserting the values of
$P_0$ and $\sigma_0$ from Eqs.(\ref{13}) and (\ref{14}) in the above
equation, it follows that
\begin{align}\nonumber
&\Phi''_0=G''_0-\frac{\alpha\sqrt{G_0}}{(1+{\delta}
R_0)}\left[-{\delta}R_0R'_0+2{\delta}(G'_0R''_0+G_0R'''_0)+2\delta
\right.\\\nonumber
&\left.\times(G'_0R'_0+G_0R'''_0)\left(\frac{2}{a_0}
+\frac{G'_0}{2G_0}\right)+2{\delta}G_0
R''_0\left(\frac{G''_0}{2G_0}-\frac{G'^2_0}
{2G^2_0}-\frac{2}{a_0^2}\right)\right.\\\nonumber
&\left.-\frac{2}{a_0}\left\{\frac{{\alpha}G'_0} {\sqrt{G_0}}
(1+2{\delta}R_0)+\frac{{\delta}R_0^2}{2}-2{\delta}G_0R''_0
-2{\delta}G_0R'_0\left(\frac{G'_0}{G_0}+\frac{1}{a_0}
\right)\right\}\right.\\\nonumber &\left.
+\frac{8{\delta}{\alpha}\sqrt{G_0}R'_0}{a_0
(1+2{\delta}R_0)^2}\right]\left[1-\frac{8{\delta}a_0R'_0}{(1
+2{\delta}R_0)}-\frac{a_0^2}{\alpha\sqrt{G_0}(1+2{\delta}R_0)}
\left\{-{\delta}R_0\right.\right.\\\nonumber
&\left.\left.{\times}R'_0+2{\delta}(G'_0R''_0+G_0R''_0)+2{\delta}
(G'_0R'_0+G_0R''_0)\left(\frac{2}{a_0} +\frac{G'_0}{2G_0}\right)
\right.\right.\\\nonumber &\left.\left.+2{\delta}G_0R'_0
\left(\frac{G''_0}{2G_0}-\frac{G'^2_0}{2G^2_0}
-\frac{2}{a_0^2}\right)+\frac{2{\alpha^2}}{a_0^2}
(1+2{\delta}R_0)\left(\frac{a_0G'_0-2G_0}{\sqrt{G_0}}
\right)\right.\right.\\\nonumber
&\left.\left.+\frac{2{\delta}R'_0}{a_0^2}(2G_0-a_0G'_0)\right\}\right]
-\frac{a_0\sqrt{G_0}\kappa}{2(1+2{\delta}R_0)}\left[\frac{2\delta}
{\alpha\kappa}\left\{-\frac{R'^2_0}{2}-\frac{R_0R''_0}{2}\right.\right.\\\nonumber
&\left.\left. +G''_0R''_0+2G'_0R'''_0+G_0R''''_0
+(G''_0R'_0+2G'_0R''_0+G_0R'''_0)\right.\right.\\\nonumber
&\left.\left.\times\left(\frac{G'_0}{2G_0}+\frac{2}{a_0}\right)
+2(G'_0R'_0+G_0R''_0)\left(\frac{G''_0}{2G_0}
-\frac{G'^2_0}{2G^2_0}-\frac{2}{a_0^2}\right)+G_0\right.\right.
\\\nonumber
&\left.\left.{\times}R'_0\left(\frac{G''_0}{2G_0}+\frac{20}{a_0^3}
+\frac{G'^3_0}{G^3_0}-\frac{3}{2}\frac{G'_0G''_0}
{G^2_0}\right)\right\}+\frac{2}{{\kappa}a_0^2}
\left\{\frac{G'_0}{\sqrt{G_0}}(1+2{\delta}R_0)\right.\right.
\\\nonumber
&\left.\left.+\frac{{\delta}R_0^2}{2\alpha}-{\delta}G_0\right.
\times\frac{2R''_0}{\alpha}
-\frac{2{\delta}G_0R'_0}{\alpha}\left(\frac{G'_0}{G_0}
+\frac{1}{a_0}\right)\right\}+\frac{1}{a_0}\left[\left\{
+\frac{2{\delta}R'_0}
{{\kappa}{\alpha}a_0^2}\right.\right.\\\nonumber
&\left.\left.\left.\times(2G_0-a_0G'_0)\frac{2(1+2{\delta}R_0)^2}
{{\kappa}a_0^2\sqrt{G_0}}\left(a_0\times G'_0-2G_0\right)\right\}
[1-\chi_0]\right.\right.\\\nonumber
&\left.\left.-\frac{1}{\alpha\kappa}
\left\{-{\delta}R_0R'_0+2{\delta}(G'_0R''_0+G_0R'''_0)
+2\delta(G'_0R'_0+G_0R'''_0)\right.\right.\right.\\\nonumber
&\left.\left.\left.\left(\frac{2}{a_0}
+\frac{G'_0}{2G_0}\right)+2{\delta}G_0
R''_0\left(\frac{G''_0}{2G_0}-\frac{2}{a_0^2}-\frac{G'^2_0}
{2G^2_0}\right)\right\}\right]+\frac{8{\delta}
\sqrt{G_0}}{a_0\kappa}\right.\\\nonumber &\left.\times\frac{1}
{(1+2{\delta}R_0)}\left(R''_0-\frac{6{\delta}R'^2_0}
{(1+2{\delta}R_0)}\right)+\frac{2{\delta}R'_0}{(1+2{\delta}R_0)^3}
\left\{\frac{1}{\alpha\kappa}\left(
-{\delta}R_0\right.\right.\right.\\\nonumber
&\left.\left.\left.{\times}R'_0+2{\delta}(G'_0R''_0+G_0R'''_0)
+2\delta(G'_0R'_0+G_0R'''_0)\left(\frac{2}{a_0}+\frac{G'_0}{2G_0}\right)
\right.\right.\right.\\\nonumber
&\left.\left.\left.+2{\delta}G_0R''_0\left(\frac{G''_0}{2G_0}
-\frac{2}{a_0^2}-\frac{G'^2_0}{2G^2_0}\right)\right)+\frac{2}
{{a_0^2\kappa}}\left(\frac{a_0G'_0-2G_0}{\sqrt{G_0}}\right)
\right.\right.\\\label{20}
&\left.\left.\times(1+2{\delta}R_0)+\frac{2{\delta}R'_0}
{a_0^2\kappa\alpha}\left(2G_0-a_0G'_0\right)\right\}\right],
\end{align}
where the subscript ``0" indicates that the quantities are evaluated
at $a=a_0$ and $\chi_0$ is given by
\begin{eqnarray}\nonumber
\chi_0=-\left[\frac{(1+2{\delta}R_0)(2G_0+a_0G'_0)
+a_0\sqrt{G_0}X_{P0}}{\sqrt{G_0}\{a_0X_{\sigma0}+4\sqrt{G_0}
(1+2{\delta}\tilde{R}_0)\}}\right].
\end{eqnarray}
For $R_0=\tilde{R}_0=constant$ and using Eq.(\ref{10}), Eq.(\ref{6})
reduces to
\begin{align}\label{21}
\sigma_0&=-\frac{4(1+2{\delta}R_0)}{a_0\kappa}\sqrt{G_0}
-\frac{{\delta}\tilde{R}_0^2}{2{\alpha}\kappa}.
\end{align}
This shows that the energy density is negative indicating the
presence of exotic matter at the throat. Moreover, Eq.(\ref{20})
turns out to be
\begin{align}\nonumber
&\Phi''_0=G''_0-2G'_0\left(\frac{3}{a_0}+\frac{a_0G'_0-2G_0}{8a_0G_0}
\right)-\left(\frac{a_0G'_0-2G_0}{a_0^2\sqrt{G_0}}\right)\\\nonumber
&\times\left[\frac{\frac{3}{2\alpha}{\delta}a_0\tilde{R}_0^2
\sqrt{G_0}+2(1+2{\delta}R_0)(4G_0+a_0G'_0)}{\frac{a_0}{2\alpha}
{\delta}\tilde{R}_0^2+4\sqrt{G_0}
(1+2{\delta}\tilde{R}_0)}\right]-\frac{\sqrt{G_0}}{a_0}
\\\label{22}
&\times\left(\frac{1}{16}+\frac{3}{2}\sqrt{G}
\right)\frac{\delta\tilde{R}_0^2}{(1+2{\delta}\tilde{R}_0)}.
\end{align}

\section{Charged Black String Thin-Shell WH}

Here, we devise thin-shell WH for the charged black string and
investigate its stability with the static background in the context
of $f(R)$ gravity. The surface pressure and energy density, under
constant Ricci scalar condition, are now obtained by using
Eqs.(\ref{3a}), (\ref{13}) and (\ref{14}) as
\begin{align}\label{24}
P_0&=\frac{({\alpha}^3a_0^3-M)(1+2{\delta}\tilde{R}_0)}{2{\pi}a_0
\sqrt{4q^2-4M\alpha{a_0+\alpha^4a^4_0}}}+\frac{{\delta}\tilde{R}^2_0}
{2\alpha},\\\label{23}
{\sigma}_0&=-\left(\frac{1+2{\delta}\tilde{R}_0}{2\pi{a^2}_0{\alpha}}\right)
\sqrt{4q^2-4M\alpha{a_0}+\alpha^4a^4_0}-\frac{{\delta}\tilde{R}^2_0}
{2\alpha}.
\end{align}
Equation (\ref{15}) leads to
\begin{align}\nonumber
&4{\alpha}^2a_0^2-\frac{4M}{a_0\alpha}+\frac{a_0{\delta}
\tilde{R}_0^2}{\alpha^2(1+2{\delta}\tilde{R}_0)}
\sqrt{4q^2-4M\alpha{a_0}+\alpha^4a^4_0}
+\frac{128a_0^2B\pi^2}{(1+2{\delta}\tilde{R}_0)}\\\label{25}&
\times\frac{
\sqrt{4q^2-4M\alpha{a_0}+\alpha^4a^4_0}}{\{-\alpha{\delta}
a_0^2\tilde{R}_0^2-8(1+2{\delta}\tilde{R}_0)\sqrt{4q^2
-4M\alpha{a_0}+\alpha^4a^4_0}\}}=0.
\end{align}
This is the required dynamical equation which the charged black
string WH threaded by exotic matter with throat radius $a_0$ must
satisfy. In this scenario, Eq.(\ref{22}) yields
\begin{align}\nonumber
&\Phi''_0=\frac{1}{Y_0}\left[\left(\frac{256q^4}{a_0^5\alpha^4}+\frac{288M^2}
{a_0^3\alpha^2}+\frac{192q^2}{a_0}-144M\alpha-\frac{576q^2M}
{\alpha^4a_0^3}\right)
\right.\\\nonumber&\left.\times(1+2{\delta}\tilde{R}_0)+\frac{2{\delta}
\tilde{R}_0^2\sqrt{4q^2-4M\alpha{a_0}+\alpha^4a^4_0}}{\alpha^3a_0^2}
\left(\frac{12q^2}{a_0\alpha}-9M\right)\right]\\\nonumber
&+\frac{{\delta}\tilde{R}_0^2}{(1+2{\delta}\tilde{R}_0)}
\left(\frac{6M}{{\alpha}a_0^2}-\frac{6M}{\alpha^2a_0^3}-\frac{3a_0\alpha^2}{2}
-\frac{\sqrt{4q^2-4M\alpha{a_0}+\alpha^4a^4_0}}{16a_0^2\alpha}\right)
\\\nonumber
& +\frac{2\alpha^2}{(4q^2-4M\alpha{a_0}+\alpha^4a^4_0)}\left[4q^2
\left(\frac{5M}{a_0^3\alpha^3}
+1-\frac{4q^2}{a_0^4\alpha^4}\right)-6Ma_0\left(\alpha\right.\right.\\\label{26}
&\left.\left. +\frac{2M}{a_0^3\alpha^2}\right)\right]+\frac{72q^2}
{a_0^4\alpha^2}-\frac{32M}{a_0^3\alpha}-10\alpha^2,
\end{align}
where
\begin{equation}\nonumber
Y_0=\frac{\sqrt{4q^2-4M\alpha{a_0}+\alpha^4a^4_0}}{\alpha}\left[\frac{a_0{\delta}
\tilde{R}_0^2}{2\alpha}+4(1+2{\delta}\tilde{R}_0)\frac{\sqrt{4q^2-4M\alpha{a_0}
+\alpha^4a^4_0}}{a_0\alpha}\right],
\end{equation}
\begin{equation}\nonumber
\tilde{R}_0=4\alpha^2\left(1+\frac{2}{a_0}\right)+\frac{16}
{{\alpha}a_0^3}\left(1+\frac{1}{a_0}\right)\left(\frac{2q^2}{{\alpha}a_0}
-M\right).
\end{equation}
\begin{figure} \centering
\epsfig{file=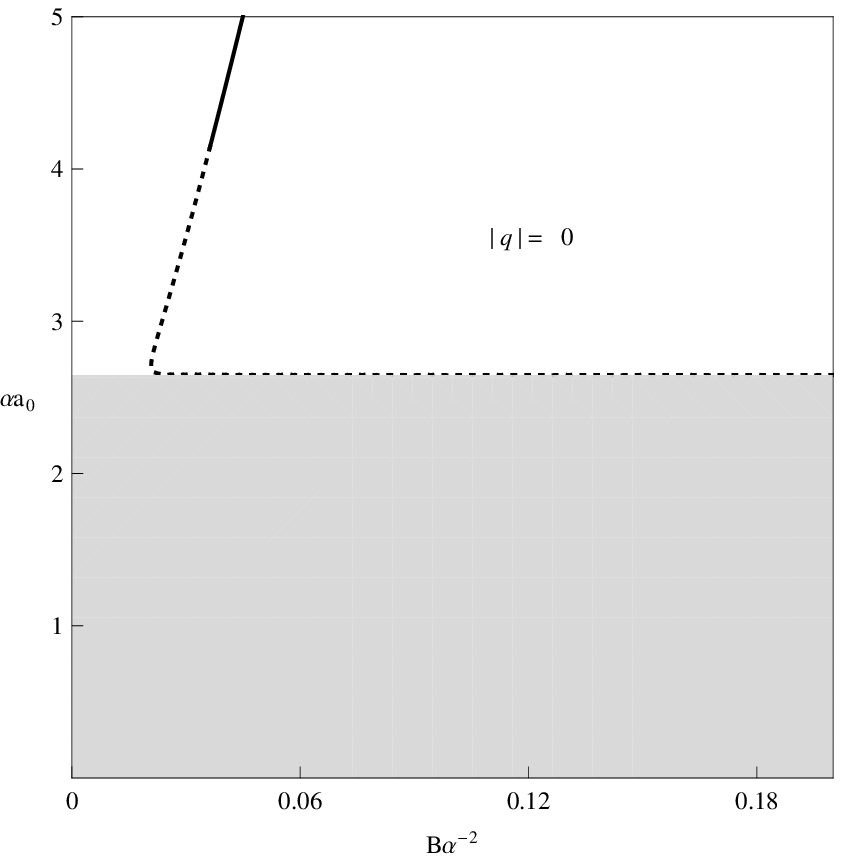,width=.42\linewidth}
\epsfig{file=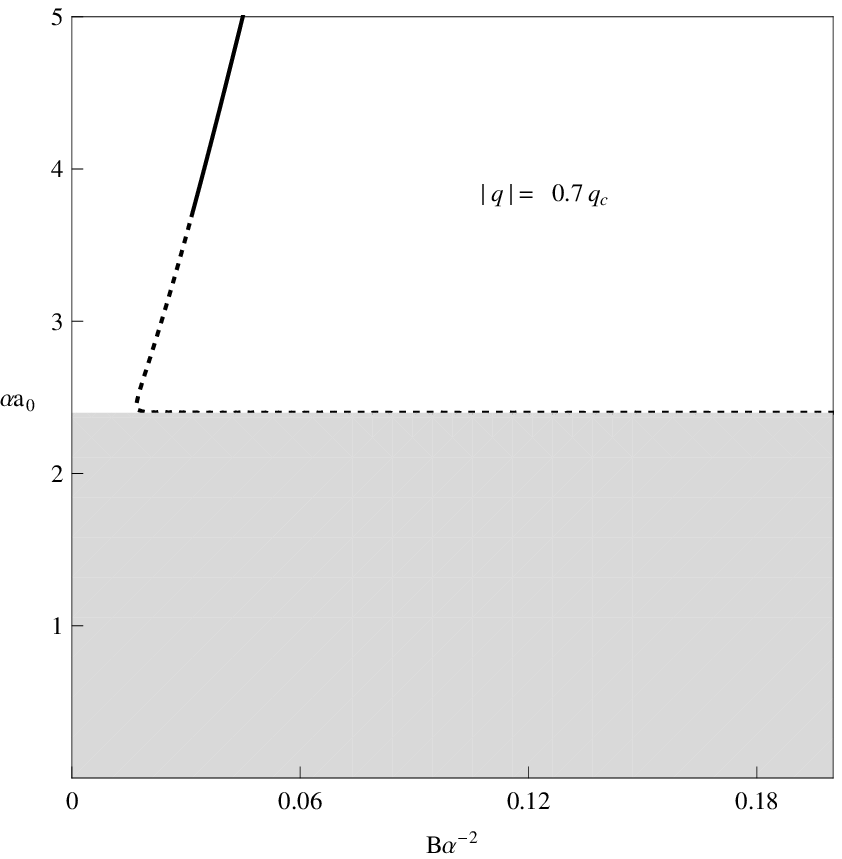,width=.42\linewidth}
\epsfig{file=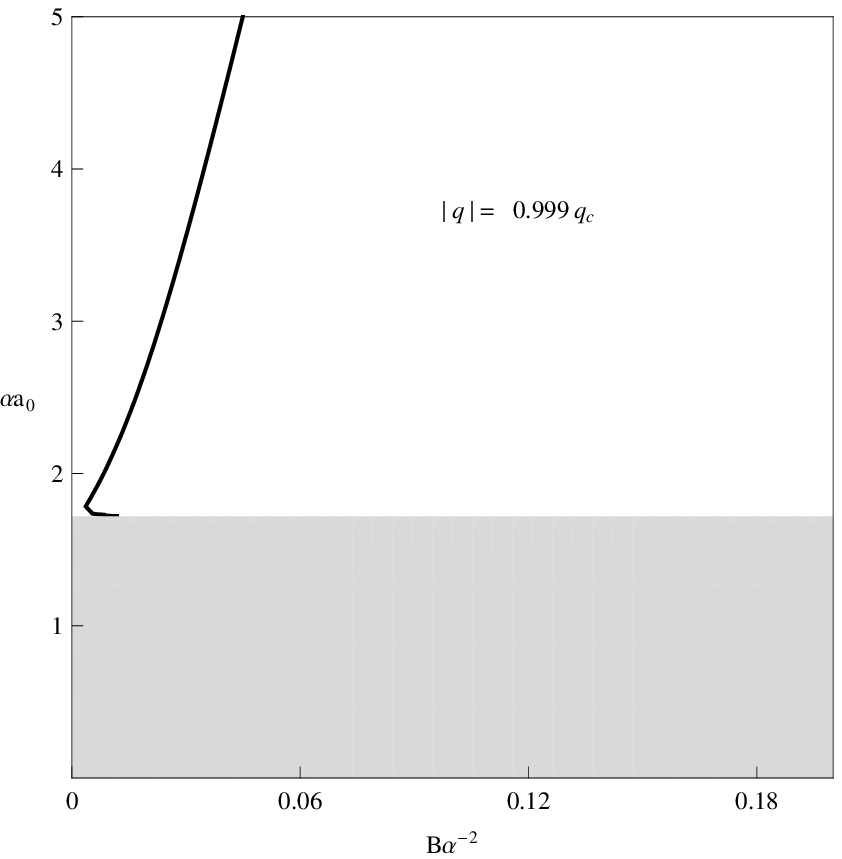,width=.42\linewidth}
\epsfig{file=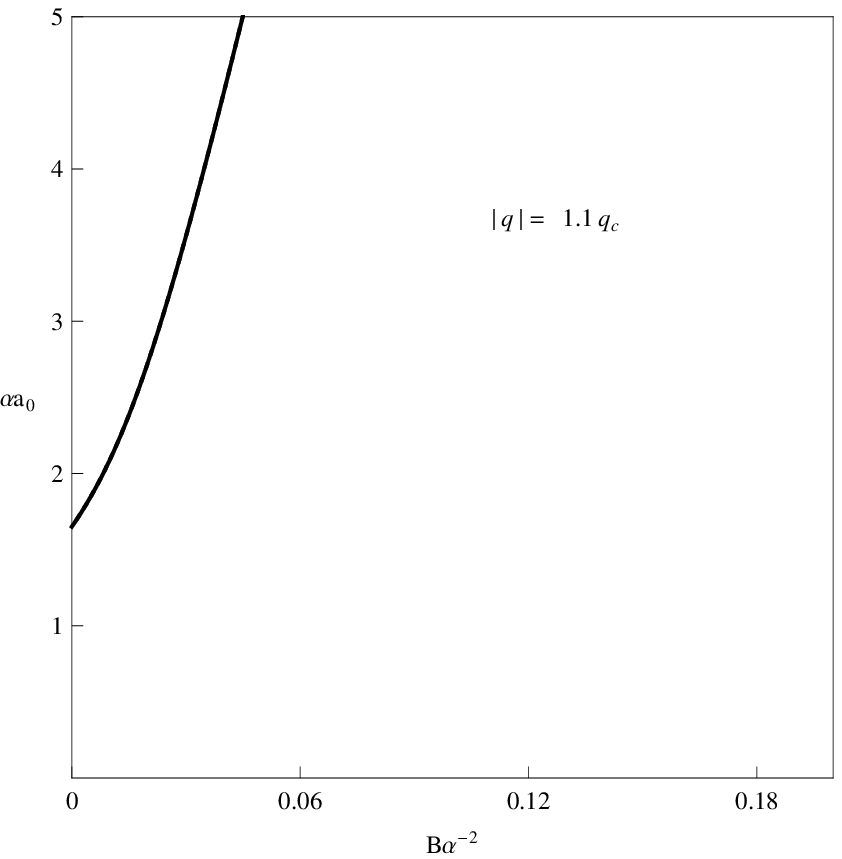,width=.42\linewidth} \caption{Wormholes of a
charged black string against radial perturbation for $M=1=\alpha$,
$\delta=0.2$ with different charge values. The dotted and solid
curves indicate unstable and stable solutions, respectively whereas
the shaded zone represents non-physical case, i.e., $a_0\leq r_h$}
\end{figure}
\begin{figure} \centering
\epsfig{file=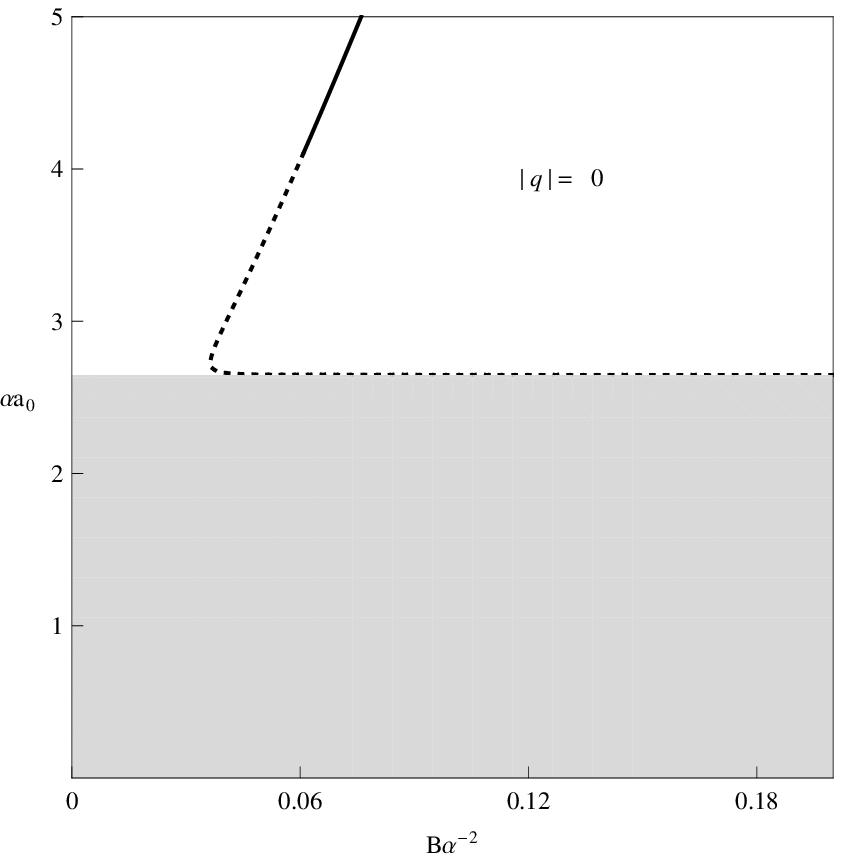,width=.42\linewidth}
\epsfig{file=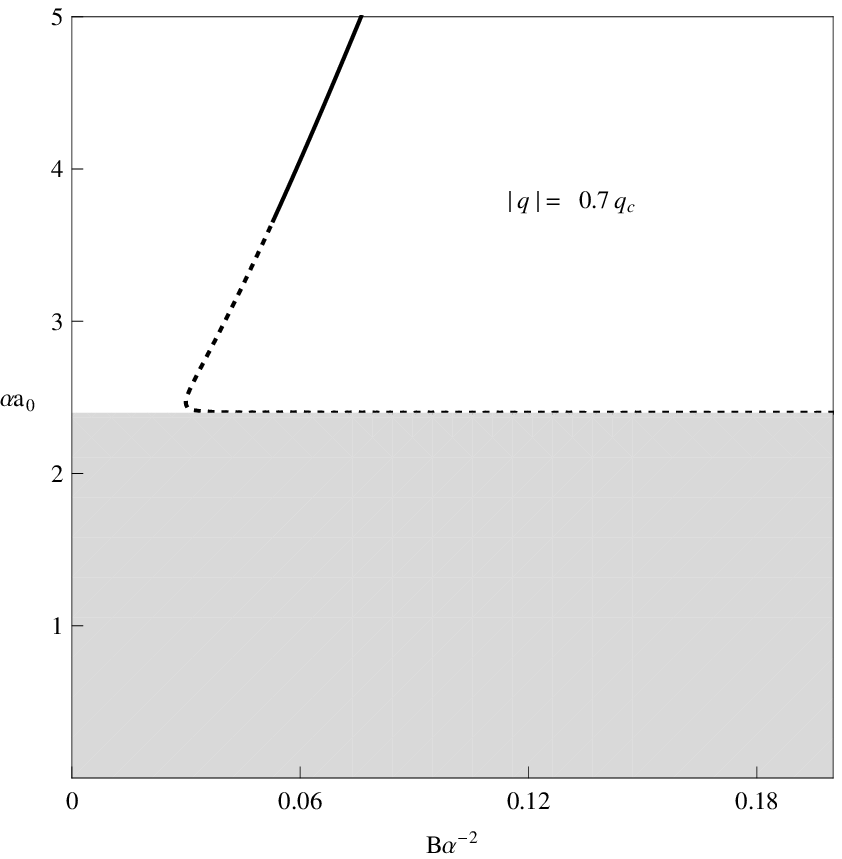,width=.42\linewidth}
\epsfig{file=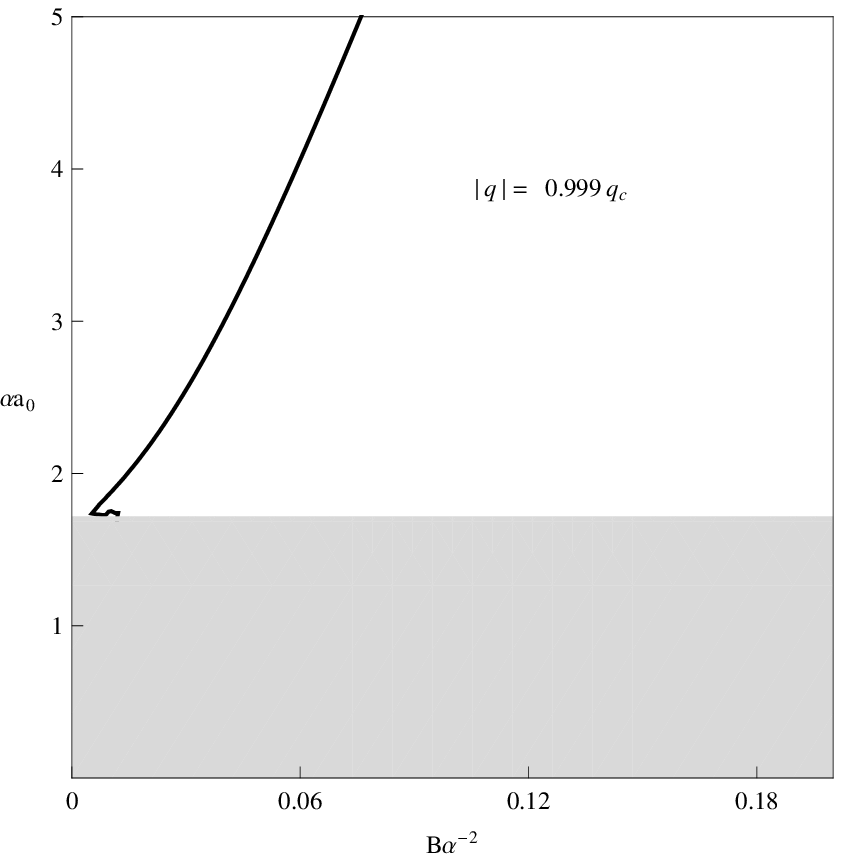,width=.42\linewidth}
\epsfig{file=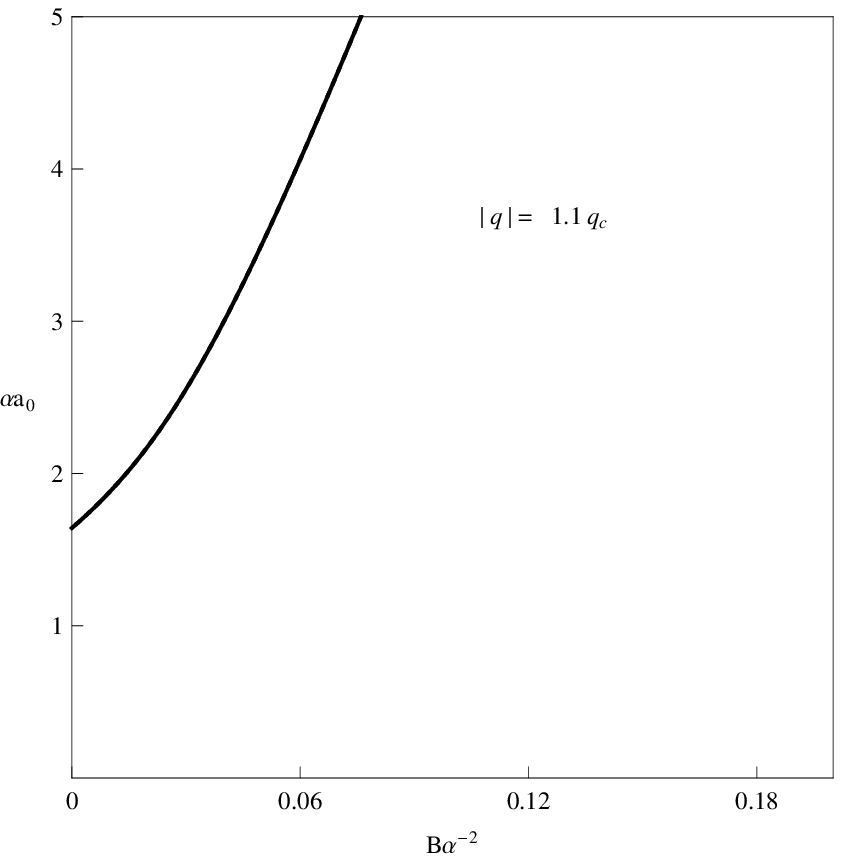,width=.42\linewidth} \caption{Wormholes of a
charged black string for $M=1=\alpha$, $\delta=0.4$ with different
values of charge.}
\end{figure}

Now, we investigate the instability and stability of the static
configurations for perturbations preserving the cylindrical symmetry
which is determined by $\Phi''<0$ or $\Phi''>0$. In all figures, the
solid line indicates the stable solution of WHs due to $\Phi''>0$
whereas $\Phi''<0$ points unstable static WH solutions which is
symbolized by dotted lines. The gray regions correspond to
non-physical zone. It is worth mentioning here that the charge
$q_c=0.866025$ determines the behavior of these solutions. This
specific value is used to construct the original metric with no
horizon.
\begin{figure} \centering
\epsfig{file=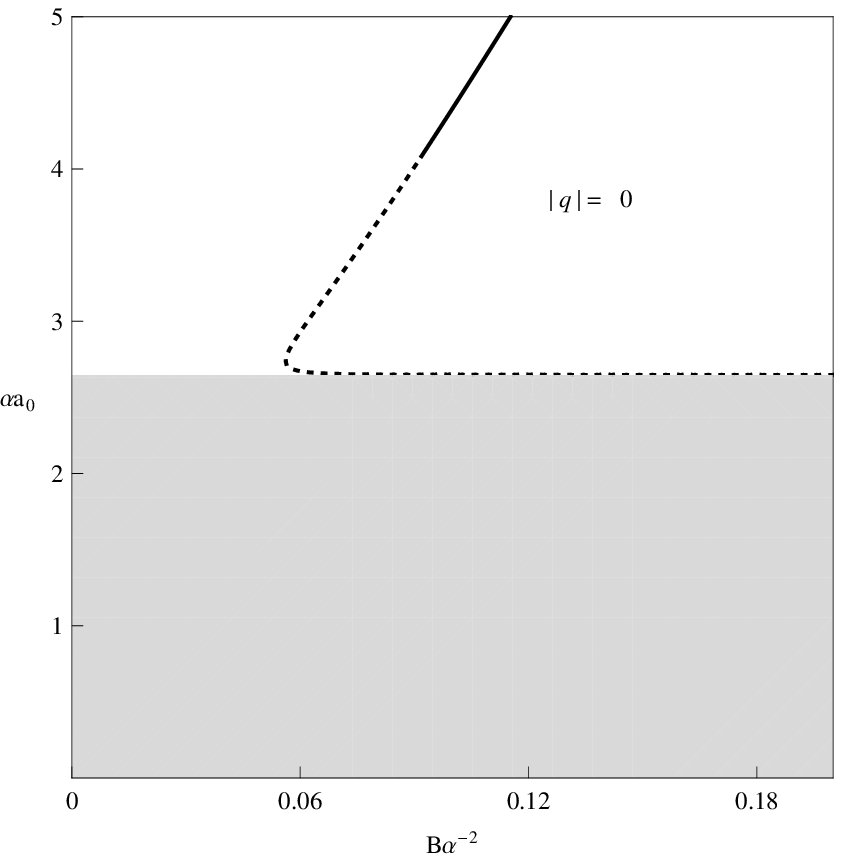,width=.42\linewidth}
\epsfig{file=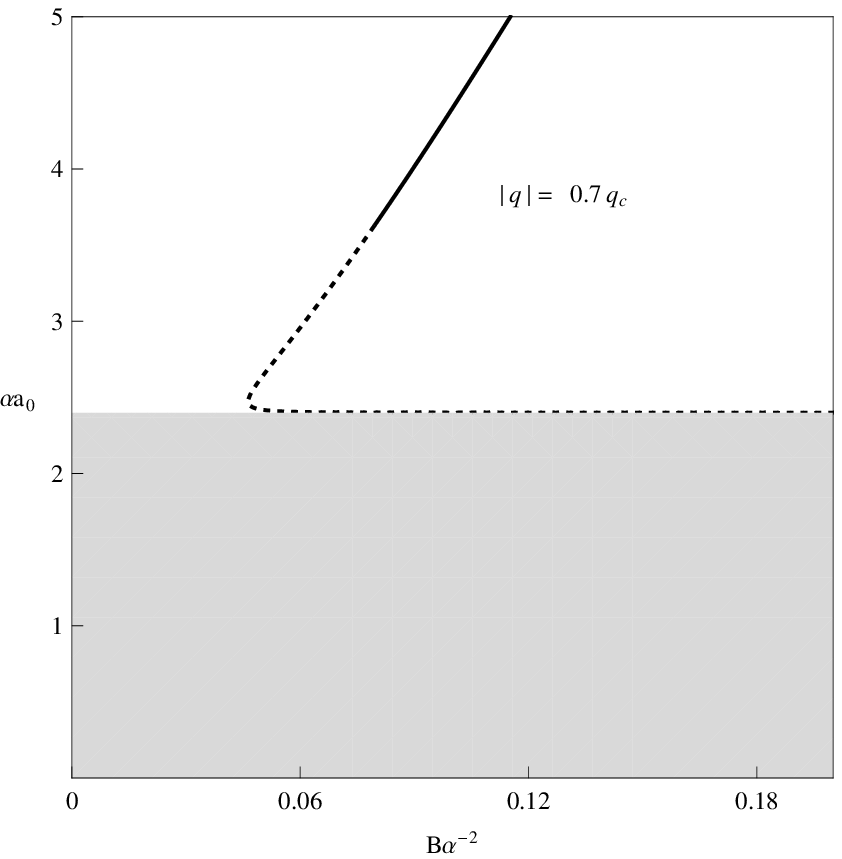,width=.42\linewidth}
\epsfig{file=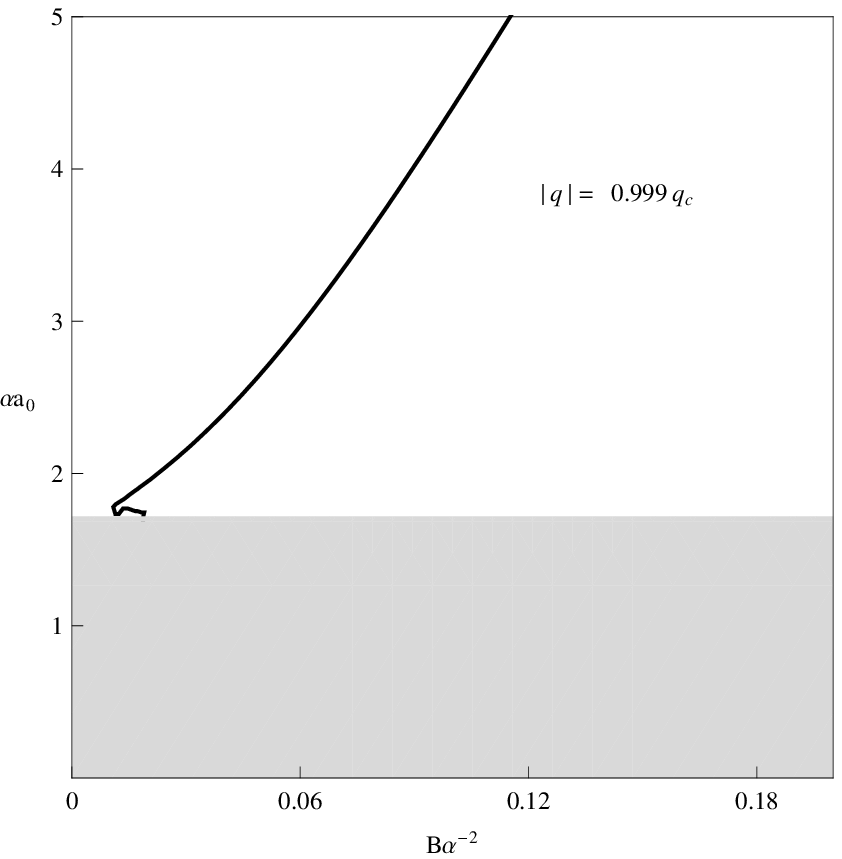,width=.42\linewidth}
\epsfig{file=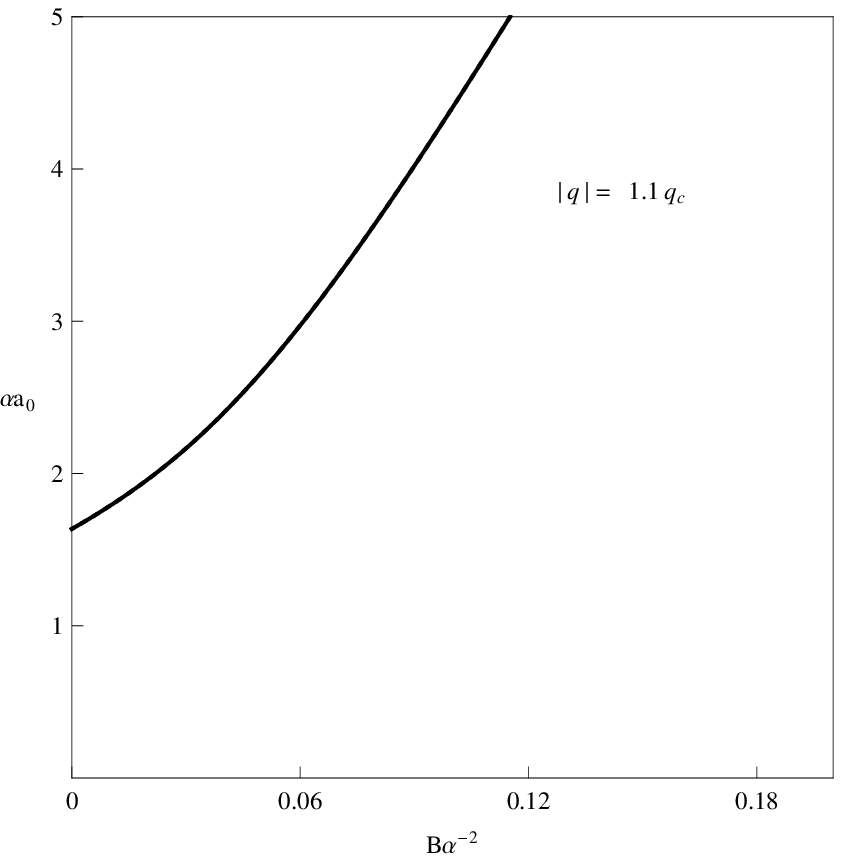,width=.42\linewidth} \caption{Wormholes of a
charged black string for $M=1=\alpha$, $\delta=0.6$ with different
values of charge.}
\end{figure}

\begin{figure} \centering
\epsfig{file=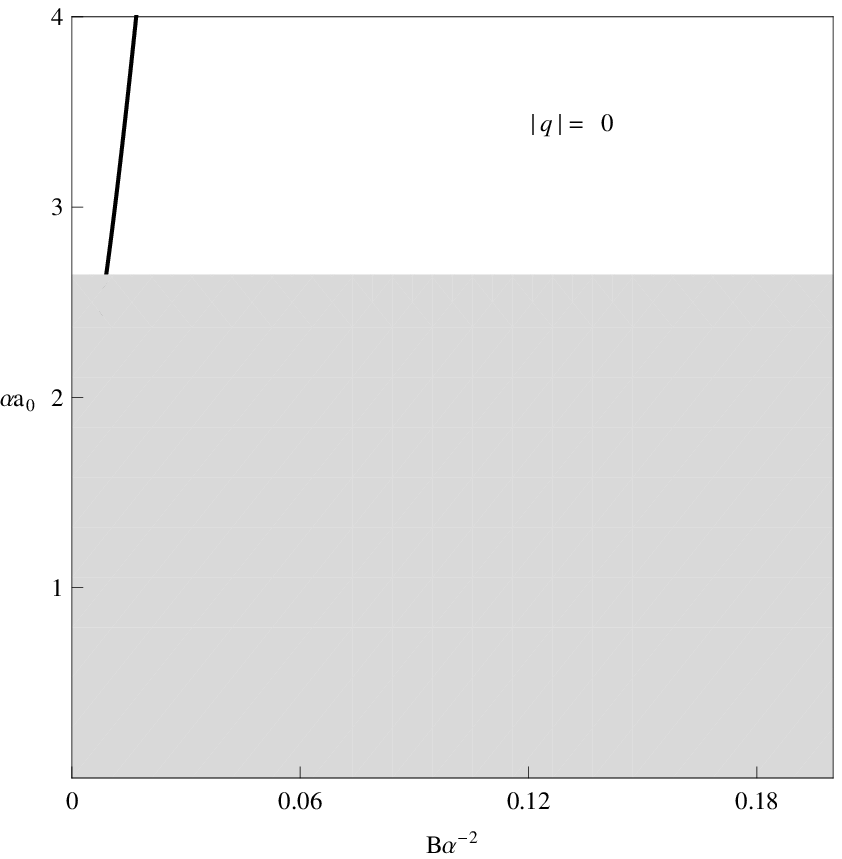,width=.42\linewidth}
\epsfig{file=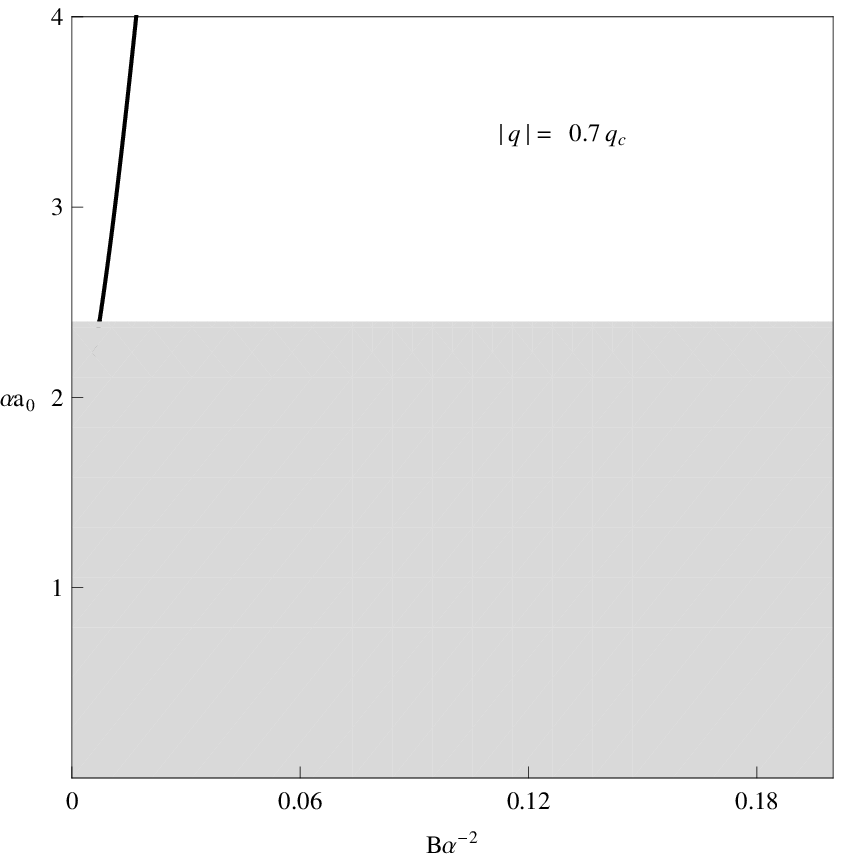,width=.42\linewidth}
\epsfig{file=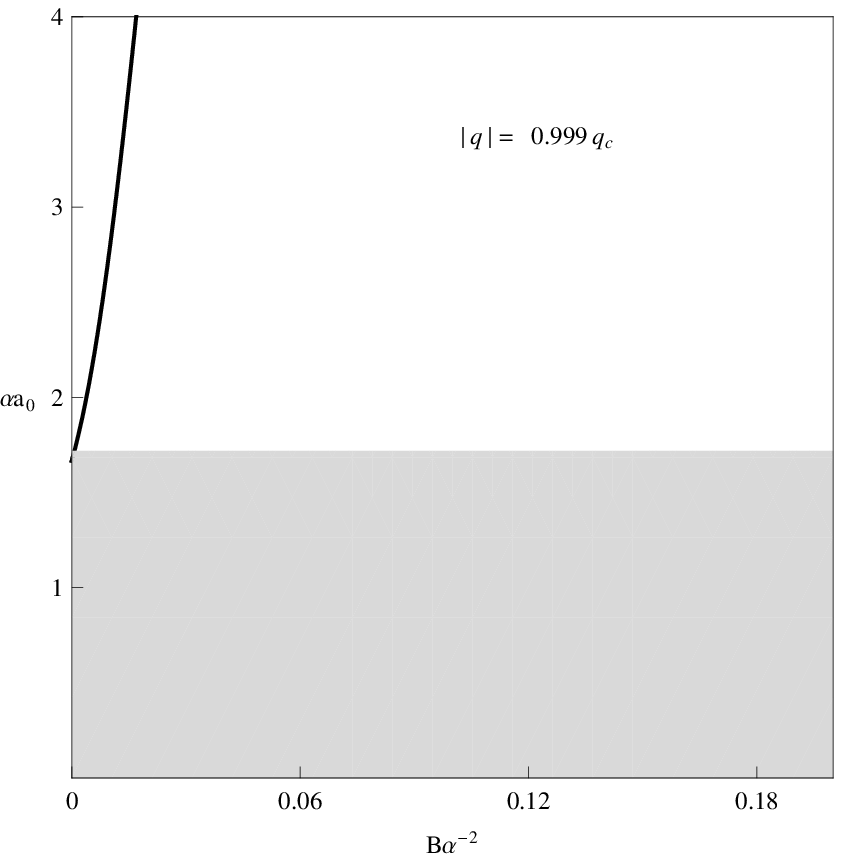,width=.42\linewidth}
\epsfig{file=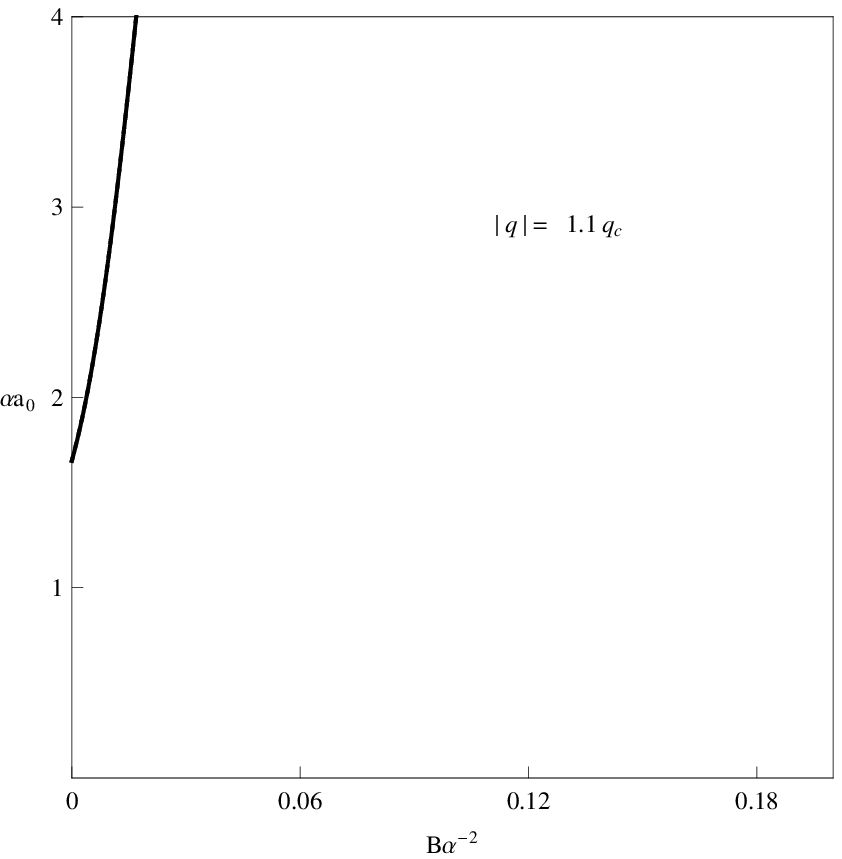,width=.42\linewidth} \caption{Wormholes of a
charged black string for $M=1=\alpha$, $\delta=0$ with different
values of charge.}
\end{figure}

In $f(R)$ model, i.e., $R+{\delta}R^2$, we take some specific values
of $\delta$ and study the stability of black string solutions.
\begin{enumerate}
\item When $\delta$~=~0.2,~0.4,~0.6.
\begin{itemize}
\item For $|q|=0$ and $|q|=0.7q_c$, i.e., $|q|$ is not very much close
to $q_c$, we find that there exist unstable and stable
configurations for some values of $B\alpha^2$ with $\delta=0.2,~0.4$
and $0.6$ as shown in Figures \textbf{1-3}. \item For $|q|\lesssim
q_c$ (Figures \textbf{1-3}), there is a stable WH static solution
for $\delta=0.2,~0.4$ and $0.6$. Further, it is seen that horizon
radius keeps on decreasing with the increase in the value of charge.
\item When $\delta=0.2,~0.4$ and $0.6$, there exist stable WH
configurations corresponding to $|q|>q_c$ as implied by Figures
\textbf{1-3}.
\end{itemize}

\item Now we make our thin-shell WH stability analysis by reducing the
equations from $f(R)$ to GR, i.e., by taking $\delta=0$.

\begin{itemize}
\item For $|q|<q_c$ and $|q|\lesssim q_c$, there always exists stable
black string WH solution for each value of $B\alpha^2$. We find from
Figure \textbf{4} that ${\alpha}a_0$ decreases upto the horizon
radius of the original manifold, and then solutions cannot be found.
We also see that increment in the charge makes the radius of the
horizon to decrease. \item There exists stable thin-shell WH
solution corresponding to $|q|>q_c$ when $\delta\rightarrow0$ as
shown in Figure \textbf{4}.
\end{itemize}
\end{enumerate}

\section{Concluding Remarks}

In this paper, we have studied the stability of WH solutions of
charged black string under perturbation with $f(R)$ terms. We have
computed the Darmois-Israel matching conditions on the matter shell. Wormholes are constructed using cut and
paste technique framed within a well-known $f(R)$ model (as a source of exotic matter). In this
scenario, dynamical equation is formulated and stability of WH
solutions (threaded by exotic matter) are investigated.

The numerical analysis is used to explore Eq.(\ref{25}) for
${\alpha}a_0$ with different values of the dark source exponent,
i.e., $\delta=0,~0.2,~0.4$ and $0.6$. The results are summarized as
follows.

\begin{enumerate}
\item Figures \textbf{1-4} indicate that the radius of the WH throat
decreases progressively till it reaches the radius of the charged
black string horizon $r_h$ for large values of $\alpha^{-2}{B}$ and
$r_h$ disappears for $|q|>q_c$. The shaded portions in the graphs
indicate regions of throat radius smaller than $r_h$.
\item It is seen that stable and unstable solutions exist for
$\delta=0.2,~0.4,~0.6$ with $|q|=0,~0.7q_c$ whereas we obtain only
stable configurations for $|q|=0.999q_c$ and $|q|=1.1q_c$ with
$\delta=0.2,~0.4,~0.6$. The radius of horizon decreases on
increasing $|q|$.
\item It is worth mentioning here that when $\delta=0$, we find stable
solutions for $|q|=0,~0.7q_c,~0.9999q_c$ and $|q|=1.1q_c$ which are
the solutions we can expect (Sharif and Azam 2013). Thus all our
results reduce to GR by taking $\delta\rightarrow0$.
\end{enumerate}

\end{document}